\documentclass[aps,prl,twocolumn,showpacs,preprintnumbers,nofootinbib,float,longbibliography,superscriptaddress]{revtex4-1}
\usepackage{epsfig}
\usepackage[dvipsnames]{xcolor}
\usepackage{float,amsmath,amssymb}
\usepackage{graphicx}
\usepackage{url}
\usepackage{bbold}
\usepackage[utf8]{inputenc}
\usepackage{physics}
\usepackage[breaklinks,colorlinks,citecolor=citcolor,urlcolor=blue,linkcolor=lcolor]{hyperref}
\definecolor{lcolor}{rgb}{0.5,0,0}
\definecolor{citcolor}{rgb}{0,0.3,0.0}

\begin{document}

\title{Anisotropic flow in fixed-target $^{208}$Pb+$^{20}$Ne collisions as a probe of quark-gluon plasma}

\author{Giuliano Giacalone}
\email{giacalone@thphys.uni-heidelberg.de}
\affiliation{Institut f\"{u}r Theoretische Physik, Universit\"{a}t Heidelberg, Philosophenweg 16, 69120 Heidelberg, Germany}

\author{Wenbin Zhao}
\email{WenbinZhao@lbl.gov}
\affiliation{Nuclear Science Division, Lawrence Berkeley National Laboratory, Berkeley, California 94720, USA}
\affiliation{Physics Department, University of California, Berkeley, California 94720, USA}

\author{Benjamin Bally}
\affiliation{ESNT, IRFU, CEA, Universit\'e Paris-Saclay, 91191 Gif-sur-Yvette, France}

\author{Shihang Shen}
\affiliation{Institute for Advanced Simulation (IAS-4), Forschungszentrum J\"ulich, D-52425 J\"ulich, Germany}

\author{\\Thomas Duguet}
\affiliation{IRFU, CEA, Universit\'e Paris-Saclay, 91191 Gif-sur-Yvette, France}
\affiliation{KU Leuven, Instituut voor Kern- en Stralingsfysica, 3001 Leuven, Belgium}

\author{Jean-Paul Ebran}
\affiliation{CEA, DAM, DIF, 91297 Arpajon, France}
\affiliation{Universit\'e Paris-Saclay, CEA, Laboratoire Mati\`ere en Conditions Extr\^emes, 91680 Bruy\`eres-le-Ch\^atel, France}

\author{Serdar Elhatisari}
\affiliation{Faculty of Natural Sciences and Engineering, Gaziantep Islam Science and Technology University, Gaziantep 27010, Turkey}

\author{Mikael Frosini}
\affiliation{CEA, DES, IRESNE, DER, SPRC, LEPh, 13115 Saint-Paul-lez-Durance, France}

\author{\\Timo A. L\"ahde}
\affiliation{Institut f\"ur Kernphysik, Institute for Advanced Simulation and J\"ulich Center for Hadron Physics, Forschungszentrum Julich, D-52425 J\"ulich, Germany}
\affiliation{Center for Advanced Simulation and Analytics (CASA), Forschungszentrum Julich, D-52425 J\"ulich, Germany}

\author{Dean Lee}
\affiliation{Facility for Rare Isotope Beams and Department of Physics and Astronomy, Michigan State University, MI 48824, USA}

\author{Bing-Nan Lu}
\affiliation{Graduate School of China Academy of Engineering Physics, Beijing 100193, China}

\author{Yuan-Zhuo Ma}
\affiliation{Facility for Rare Isotope Beams and Department of Physics and Astronomy, Michigan State University, MI 48824, USA}

\author{Ulf-G. Mei\ss ner}
\affiliation{Helmholtz-Institut f\"ur Strahlen- und Kernphysik, Universit\"at Bonn, D-53115 Bonn, Germany}
\affiliation{Bethe Center for Theoretical Physics, Universit\"at Bonn, D-53115 Bonn, Germany}
\affiliation{Institute for Advanced Simulation (IAS-4), Forschungszentrum J\"ulich, D-52425 J\"ulich, Germany}
%\affiliation{Tbilisi State University, 0186 Tbilisi, Georgia}

\author{\\Govert Nijs}
\affiliation{Theoretical Physics Department, CERN, CH-1211 Gen\`eve 23, Switzerland}

\author{Jacquelyn Noronha-Hostler}
\affiliation{Illinois Center for Advanced Studies of the Universe, Department of Physics, University of Illinois at
Urbana-Champaign, Urbana, IL 61801, USA}

\author{Christopher Plumberg}
\affiliation{Natural Science Division, Pepperdine University, Malibu, CA 90263, USA}

\author{Tom\'as R.\ Rodr\'iguez}
\affiliation{Departamento de Estructura de la Materia, F\'isica T\'ermica y Electr\'onica and IPARCOS, Universidad Complutense de Madrid, E-28040 Madrid, Spain}

\author{\\Robert Roth}
\affiliation{Institut f\"ur Kernphysik, Technische Universit\"at Darmstadt, 64289 Darmstadt, Germany}
\affiliation{Helmholtz Forschungsakademie Hessen f\"ur FAIR, GSI Helmholtzzentrum, 64289 Darmstadt, Germany}

\author{Wilke van der Schee}
\affiliation{Theoretical Physics Department, CERN, CH-1211 Gen\`eve 23, Switzerland}
\affiliation{Institute for Theoretical Physics, Utrecht University, 3584 CC Utrecht, The Netherlands}
\affiliation{NIKHEF, Amsterdam, The Netherlands}

\author{Bj\"orn Schenke}
\email{bschenke@bnl.gov}
\affiliation{Physics Department, Brookhaven National Laboratory, Upton, NY 11973, USA}

\author{Chun Shen}
\email{chunshen@wayne.edu}
\affiliation{Department of Physics and Astronomy, Wayne State University, Detroit, Michigan 48201, USA}
\affiliation{RIKEN BNL Research Center, Brookhaven National Laboratory, Upton, NY 11973, USA}

\author{Vittorio Som\`a}
\affiliation{IRFU, CEA, Universit\'e Paris-Saclay, 91191 Gif-sur-Yvette, France}

\begin{abstract}
The System for Measuring Overlap with Gas (SMOG2) at the LHCb detector enables the study of fixed-target ion-ion collisions at relativistic energies ($\sqrt{s_{\rm NN}}\sim100$ GeV in the centre-of-mass). With input from \textit{ab initio} calculations of the structure of $^{16}$O and $^{20}$Ne, we compute 3+1D hydrodynamic predictions for the anisotropic flow of Pb+Ne and Pb+O collisions, to be tested with upcoming LHCb data. This will allow the detailed study of quark-gluon plasma (QGP) formation as well as experimental tests of the predicted nuclear shapes. Elliptic flow ($v_2$) in Pb+Ne collisions is greatly enhanced compared to the Pb+O baseline due to the shape of $^{20}$Ne, which is deformed in a bowling-pin geometry. Owing to the large $^{208}$Pb radius, this effect is seen in a broad centrality range, a unique feature of this collision configuration. Larger elliptic flow further enhances the quadrangular flow ($v_4$) of Pb+Ne collisions via non-linear coupling, and impacts the sign of the kurtosis of the elliptic flow vector distribution ($c_2\{4\}$). Exploiting the shape of $^{20}$Ne proves thus an ideal method to investigate the formation of QGP in fixed-target experiments at LHCb, and demonstrates the power of SMOG2 as a tool to image nuclear ground states.
\end{abstract}

\preprint{CERN-TH-2024-074}

\maketitle

%%%%%%%%%%%%%%%%%%%%%%%%%%%%%%%%%%%%%%%%%%%%%%%%%%%%%%%%%%%%

%%%%%%%
%%%%%%%
%\noindent {\it 1. Introduction. }
%%%%%%%
%%%%%%%

Over the past decades, high-energy collisions of atomic nuclei have provided significant insights into the phase diagram of strong-interaction matter across a wide range of temperatures and densities \cite{Busza:2018rrf,Paquet:2023rfd}. These investigations also offer a new way to study nuclear structure, allowing a direct view of many-body correlations of nucleons (e.g., deformations) in the ground states of the colliding nuclei, overcoming some limitations of low-energy nuclear structure experiments \cite{STAR:2024wgy}.

At ultra-relativistic energies, experiments at the Relativistic Heavy Ion Collider (RHIC) and CERN Large Hadron Collider (LHC) have revealed the formation of quark-gluon plasma (QGP) \cite{Harris:2024aov}, the hot phase of nuclear matter where quarks and gluons are no longer confined inside protons and neutrons. The formation of QGP has been confirmed through the combined observation of anisotropic flow \cite{Shen:2020mgh,Ollitrault:2023wjk}, which characterizes the bulk dynamics of the matter, and of the modifications of the hard probes, energetic particles interacting with the hot plasma \cite{Apolinario:2022vzg}. These effects have been observed in large systems, like Au+Au and Pb+Pb collisions. In smaller systems, like proton-proton (p+p) and proton-nucleus (p+Pb) collisions, anisotropic flow is observed ~\cite{PHENIX:2018lia, ATLAS:2014qaj,CMS:2016fnw,ATLAS:2018ngv,ALICE:2019zfl,ATLAS:2021jhn,Zhao:2022ayk,STAR:2022pfn,Chen:2023njr} albeit in the absence of hard-probe modifications \cite{ALICE:2016dei,CMS:2016xef,ATLAS:2022iyq}, leaving the question of small-sized QGP formation open \cite{Nagle:2018nvi,Schenke:2021mxx,Noronha:2024dtq}.
 
Fresh insights are expected from the LHC’s new fixed-target program. The LHCb detector hosts the System for Measuring Overlap with Gas (SMOG2) \cite{LHCb:2021ysy,LHCb:2022qvj,Mariani:2022klh}, enabling beam-target collisions at relativistic energy in the center-of-mass frame ($\sqrt{s_{\rm NN}}=68.5$ GeV for beam nucleons boosted to 2.5 TeV). Intermediate-sized systems like lead-neon (Pb+Ne) and lead-oxygen (Pb+O) collisions provide a critical middle ground to understanding the transition from a large to a potentially small QGP. Additionally, due to the lower energies compared to, e.g., symmetric O+O collisions, these systems may present higher baryon density, allowing further tests of the equation of state for baryon-rich nuclear matter \cite{Annala:2019puf,OmanaKuttan:2022aml,Giacalone:2023cet,Shen:2023awv} in system sizes that have not been explored at the RHIC Beam Energy Scan (BES) or elsewhere, which focused on Au+Au collisions \cite{HADES:2019auv,STAR:2021yiu,HADES:2022osk}.

Most importantly, light-heavy ion collisions help reduce uncertainties related to the initial state geometry. Our understanding of the interaction geometry of proton-nucleus, deuteron-nucleus, and $^3$He-nucleus collisions suffers from a poor knowledge of the internal spatial structure of nucleons at high energy \cite{Schenke:2021mxx}. With nuclei like $^{20}$Ne and $^{16}$O, the initial state geometry will be dominated by the nucleon distributions, and we can rely on precise nuclear structure calculations based on effective field theories of low-energy QCD. This leads to a clearer understanding of the initial conditions of the collisions, which is crucial for interpreting the dynamics of the formed matter. In turn, this also allows new experimental tests of the predicted nuclear shapes \cite{Giacalone:2024luz} and alpha-clustering \cite{YuanyuanWang:2024sgp,Zhang:2024vkh} in the ground states of these isotopes.

Characterizations of the matter produced in fixed-target experiments have only just begun. The LHCb collaboration has recently published the measurement of $J/\Psi$ and $D_0$ production in Pb+Ne collisions \cite{LHCb:2022qvj}.  The resulting ratio between the two cross sections is consistent with the $p$+Ne baseline, even in high-multiplicity events, suggesting that hot-medium interactions may not be effective. Therefore, it is crucial to explore if sizable anisotropic flow is observed in these collisions. Experimental studies of the soft sector and the collective flow have not been performed yet, while, surprisingly, theoretical studies are also missing. To motivate and underscore the potential of the experimental effort, in this Letter we aim to fill this important gap. 

%%%%%%%
%%%%%%%
%\noindent {\it 2. Model and set-ups. Figure 1 }
%%%%%%%
%%%%%%%

Owing to recent advances in the understanding of the ground-state structure of $^{16}$O and $^{20}$Ne, as well as of the longitudinal structure of heavy-ion collisions, we work within an end-to-end hydrodynamic framework based on PGCM/NLEFT+3D-Glauber+MUSIC+UrQMD simulations of Pb+Ne and Pb+O collisions. 

The simulations start with configurations of nucleons for $^{20}$Ne and $^{16}$O obtained from either \textit{ab initio} Projected Generator Coordinate Method (PGCM) \cite{Frosini:2021fjf,Frosini:2021sxj,Frosini:2021ddm} or Nuclear Lattice Effective Field Theory (NLEFT) \cite{Lee:2008fa,Lahde:2019npb,Lee:2020meg} calculations, as presented in Ref.~\cite{Giacalone:2024luz}.  The PGCM nucleons are sampled either independently from the intrinsic nuclear densities (labeled hereafter \textit{Independent} configurations) or by enforcing two protons and two neutrons to sit close to $\alpha$-cluster centers (labeled \textit{Cluster} configurations).  For the NLEFT calculation, we do not take into consideration the moderate sign problem that affects the Monte Carlo sampling, and construct expectation values in the final state by assigning a positive weight to each event, which is done by considering only positive-sign nuclear configurations in the hydro calculations (see \cite{Giacalone:2024luz} for a detailed discussion of this matter). This approximation is good enough for the qualitative nature of the present study. For $^{208}$Pb, we assume independent nucleons within a matter distribution, $\rho(r)$, corresponding to the measured charge density in Woods-Saxon (W-S) form:
\begin{equation}
\label{eq:WS}
    \rho(r) \propto \left ( 1 + \exp  [ (r-R) / a  ]  \right )^{-1},
\end{equation}
with $R=6.62$ fm, and $a=0.55$ fm \cite{DeVries:1987atn}. 

Details on the three-dimensional energy- and baryon-density deposition model within the Glauber picture \cite{Shen:2022oyg} are provided in the Supplemental Material (SM). The MUSIC code \cite{Schenke:2010nt,Paquet:2015lta,Denicol:2018wdp} solves 3+1D relativistic hydrodynamic equations by evolving the initialized densities with lattice QCD based equation of state~\cite{Monnai:2019hkn}. When the local energy density drops below 0.45 GeV$/{\rm fm^{3}}$, we switch to a hadron gas phase described within the hadron cascade model UrQMD \cite{Bass:1998ca,Bleicher:1999xi}. For each nuclear structure scenario, we run 200k minimum bias events for both Pb+Ne and Pb+O collisions. For each collision event, the particlization to hadrons and the UrQMD evolution are repeated, so that the event is oversampled until we obtain $10^6$ hadrons per hydro-surface, irrespective of centrality. 

Final state observables are evaluated using all oversampled UrQMD events, which effectively erases \textit{non-flow} contributions and statistical fluctuations. The model parameters used in the calculations are the same as in Ref.~\cite{Zhao:2022ugy}. They lead to a good description of the pseudorapidity ($\eta$) dependence of the charged yields in $p/d$/$^{3}$He+Au collisions at top RHIC energy.  The simulations are performed in the centre-of-mass frame, where $\sqrt{s_{\rm NN}}=68.5$ GeV. The final-state hadrons are then boosted by $\Delta y = y_{\rm beam}({\rm 5.02~TeV}) -y_{\rm beam}({\rm 68.5~GeV})$ to the lab frame.  
The event centrality is determined from the distribution of the charged hadron multiplicity within the acceptance of the LHCb spectrometer, $2<\eta<5$, with 0\% corresponding to the high-multiplicity limit.

\begin{figure}[t]
      \includegraphics[width=0.8\columnwidth]{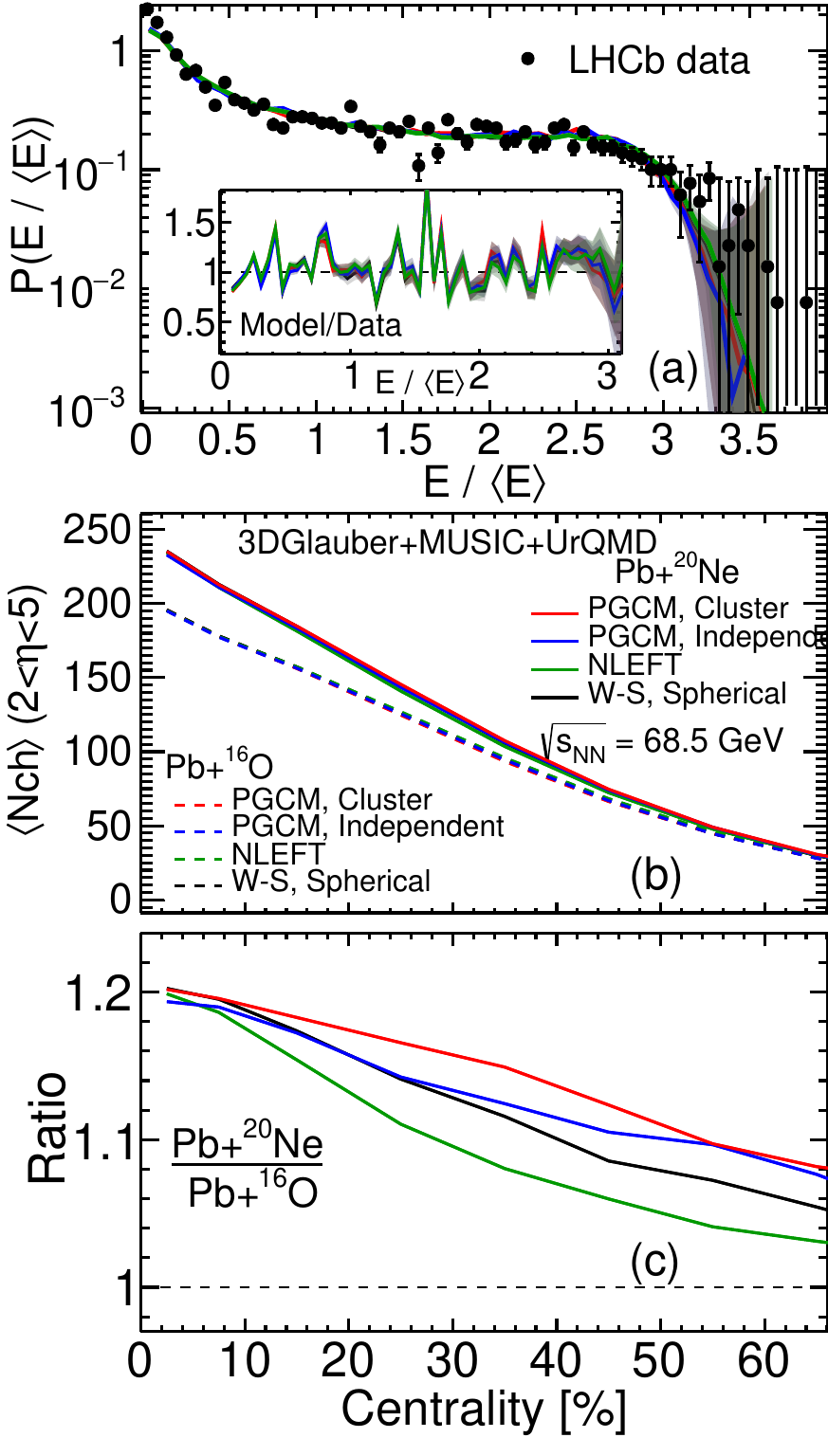}
    \caption{(a): Probability density of the rescaled energy, $E/\langle E \rangle$, in  Pb+Ne and Pb+O collisions at $\sqrt{s_{\rm NN}}=$ 68.5 GeV. Lines are hydrodynamic predictions for different nuclear structure inputs. Symbols are experimental data from the LHCb Collaboration \cite{LHCb:2021ysy}. (b): Hydrodynamic prediction for the charged multiplicity in the LHCb acceptance as a function of the collision centrality.  We plot as well the ratio of charged multiplicity taken between Pb+Ne and Pb+O collisions in panel (c). Errors are statistical only and of the same size as the shown lines when not vibisle. } 
    \label{fig:pE}
\end{figure}

%% Figure 1
To start,  we calculate the total energy, $E$, of the final-state hadrons in the lab frame. In Fig.~\ref{fig:pE}(a), the minimum-bias distribution of $E$, rescaled by the mean value of the full sample, is compared to the histogram of the rescaled energy collected in the electromagnetic calorimeter of LHCb in Pb+Ne collisions \cite{LHCb:2021ysy}, which should be strongly correlated with the total event $E$. Excellent agreement between data and our 3+1D model result is found. We emphasize that agreement is as good as that obtained by the LHCb collaboration via a dedicated Glauber fit of the energy histogram (Fig.~18 in \cite{LHCb:2021ysy}). 

Next, we predict the centrality dependence of the average charged multiplicity, $\langle N_{\rm ch} \rangle$. In Fig.~\ref{fig:pE}(b)-(c), it displays a steeper trend as a function of centrality in Pb+Ne collision than in Pb+O collisions. We understand this as follows. Due to the larger $^{20}$Ne size, Pb+Ne collisions have a higher nucleus-nucleus cross section, which implies larger impact parameters at the same centrality percentile. The size of the $^{208}$Pb nucleus is however the same in both systems. Therefore, the number of nucleons   that do not hit the incoming $^{208}$Pb increases more rapidly with centrality for a $^{20}$Ne target than for a $^{16}$O target, explaining the steeper trend in Figs.~\ref{fig:pE}(b)-(c). 

The qualitative behavior is robust against variations in the nuclear structure input. The NLEFT curve has the steepest decrease because it predicts a larger Ne/O ratio for the nuclear radius than the PGCM calculations \cite{Giacalone:2024luz}. For completeness, we repeat our calculation with spherical $^{20}$Ne and $^{16}$O nuclei parametrized via Eq.~(\ref{eq:WS}) with $R=$ 2.8 fm, $a=$ 0.57 fm for $^{20}$Ne, and $R=$ 2.61 fm, $a=$ 0.51 fm for $^{16}$O  \cite{DeVries:1987atn}. This leads to essentially the same prediction in Fig.~\ref{fig:pE}(c), confirming that the trend is driven by a size effect rather than the $^{20}$Ne deformation.

%\noindent {\it 3. Collision geometry. Figure 2}

Moving to results involving the anisotropy of the overlap area in Pb+Ne collisions,  the novelty and uniqueness of this configuration can be grasped by analyzing the initial-state ellipticity, $\varepsilon_2$, which sources the final-state elliptic flow. This is quantified via the quadrupole moment of the energy-density field, $\tau e (r,\phi_r)$ [GeV/fm$^2$], at the beginning of hydrodynamics \cite{Teaney:2010vd}:
\begin{equation}
 \varepsilon_2 = \frac{\left | \int  rdrd\phi_r ~ r^2 e^{i2\phi_r} \tau e (r,\phi_r) \right | }{\int rdrd\phi_r ~r^2 \tau e (r,\phi_r)}, \hspace{20pt} \varepsilon_2\{2\}^2 \equiv \langle \varepsilon_2^2 \rangle ,
\end{equation}
where the average is over events at a given centrality.  

\begin{figure}[t]
    \includegraphics[width=0.85\columnwidth]{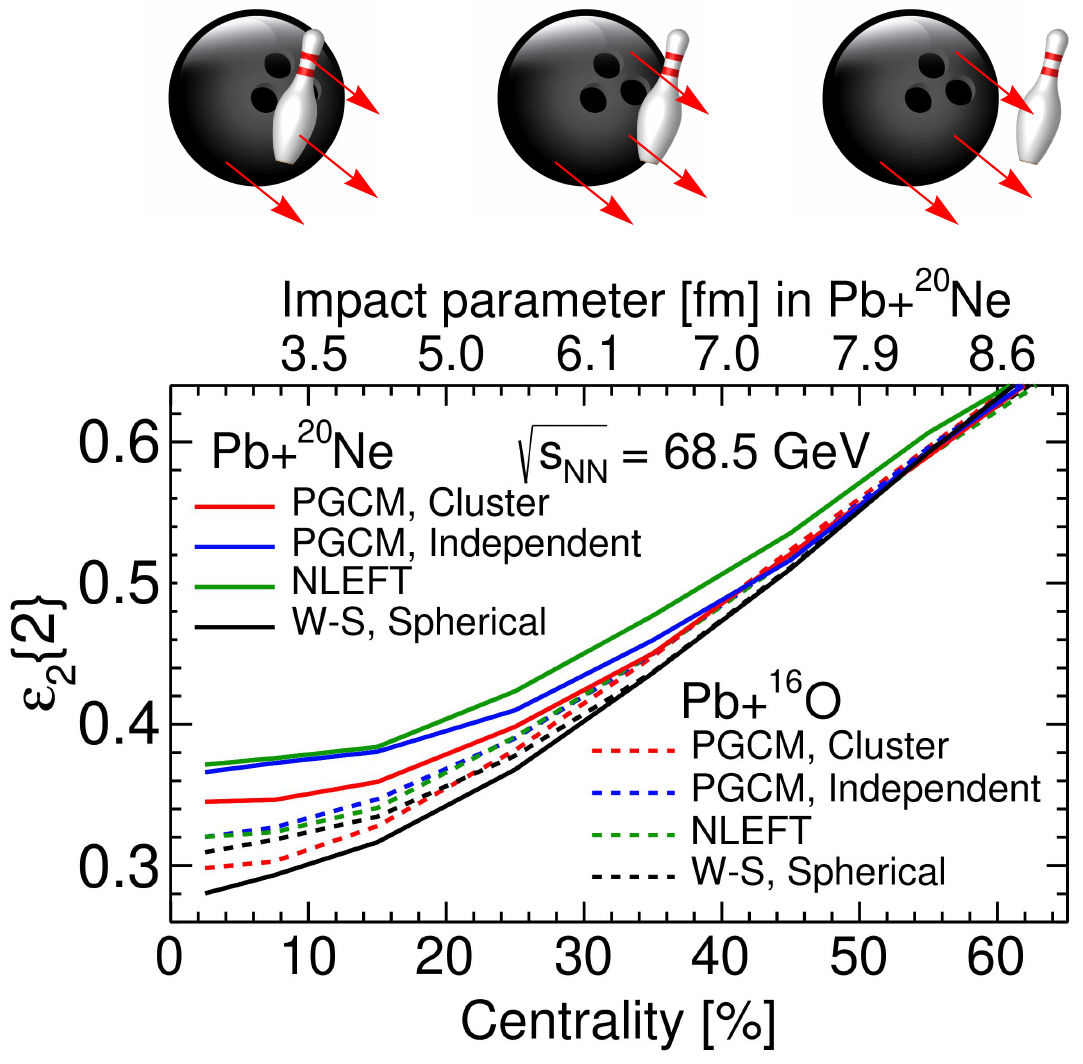}
    \caption{Playing ultra-relativistic bowling at the LHC. The rms initial-state ellipticity, $ \varepsilon_2\{2\}$ is plotted as a function of the centrality percentile in Pb+Ne collisions (solid lines) and Pb+O collisions (dashed lines) at $\sqrt{s_{\rm NN}}=$ 68.5 GeV, for various nuclear structure models. Errors are statistical only and of the same size as the line width when not visible. The sketches on top of the figure illustrate the collision geometry based on the average impact parameters of the Pb+Ne collisions (labeling the upper axis of the plot). We recall  the $^{208}$Pb W-S radius is around 6.6 fm, and about 2.8 fm for $^{20}$Ne  \cite{DeVries:1987atn}.}
    \label{fig:e2}
\end{figure}

Due to the peculiar shape of $^{20}$Ne,  deformed into a bowling-pin-like $^{16}$O+$\alpha$ configuration \cite{Giacalone:2024luz},  performing Pb+Ne collisions amounts to playing bowling with the ball (lead) thrown towards the bowling pin (neon) target. When the hit neon lies fully within the area of the lead nucleus, its entire shape is resolved. Naturally, we expect the eccentricity of the overlap region to be nearly constant for those impact parameters that correspond to such a configuration. Given the large size of $^{208}$Pb, this explains  why the variation of $\varepsilon_2\{2\}$ observed in Fig.~\ref{fig:e2} in Pb+Ne collisions is so small up to impact parameters of order $b\sim5$ fm. This   behavior seems solid and genuinely induced by the large quadrupole deformation of $^{20}$Ne:
the calculation for a spherical W-S $^{20}$Ne nucleus leads   to a steeper (though, predictably, still mild compared to a symmetric, e.g., oxygen-oxygen configuration) centrality dependence for $\varepsilon_2\{2\}$, consistent with the trend of Pb+O collisions for both deformed and spherical $^{16}$O nuclei. In terms of observable quantities, we expect thus an enhancement of the elliptic flow of Pb+Ne collisions relative to Pb+O collisions  in a broad centrality range.

%\noindent {\it 3. Anisotropic flow. Figure 3 + 4}

With this insight in mind we look at the anisotropic flow in momentum space, defined by the set of Fourier harmonics that characterize the azimuthal angle $\phi$ dependence of the charged hadron spectrum:
\begin{equation}
    \frac{dN_{\rm ch}}{d\phi} \propto 1 + 2 \sum_{n = 1}^\infty v_n \cos [n(\phi-\phi_n)],
\end{equation}
where $v_n$ is the magnitude of the $n$-th order harmonic. At a given centrality, we evaluate the root mean square value of the distribution of the coefficient (see the SM for further derivations),
\begin{equation}
    v_n\{2\} = \sqrt{\langle v_n^2 \rangle}.
\end{equation}

\begin{figure*}
      \includegraphics[width=.9\linewidth]{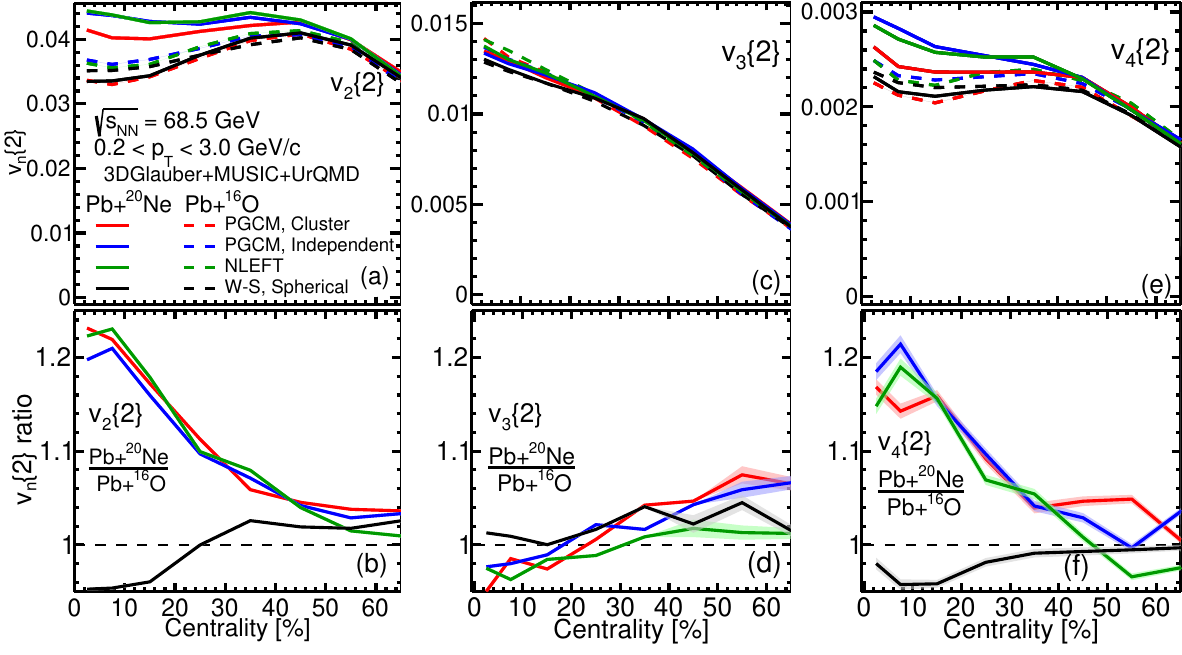}
    \caption{Anisotropic flow coefficients in Pb+Ne collisions (solid lines) and Pb+O collisions (dashed lines) at $\sqrt{s_{\rm NN}}=$ 68.5 GeV, calculated within the LHCb acceptance, as a function of the collision centrality. Upper panels show predictions for the rms elliptic flow, $v_2\{2\}$ (a), triangular flow,  $v_3\{2\}$ (c), and quadrangular flow, $v_4\{2\}$ (e). Lower panels show the corresponding ratios taken between  Pb+Ne and Pb+O collisions. Different line colors correspond to different nuclear structure inputs. Errors are statistical only and of the same size as the shown lines when not visible.}
    \label{fig:vnratio}
\end{figure*}

%% Figure 3
Figure~\ref{fig:vnratio}(a) displays our predictions for the elliptic flow ($n=2$) of Pb+Ne and Pb+O collisions. The centrality dependence of $v_2\{2\}$ in Fig.~\ref{fig:vnratio}(a) is rather flat, especially when a deformed $^{20}$Ne is considered. Computing the Pb+Ne/Pb+O ratio, shown in Fig.~\ref{fig:vnratio}(b), highlights instead the strong impact of the shape of $^{20}$Ne, which enhances the elliptic flow in central Pb+Ne collisions by more than 20\%. The signal survives up to large centralities, confirming the intuition from Fig.~\ref{fig:e2}. This showcases the unique power of SMOG2, and in general of asymmetric Pb+X collisions \cite{Bally:2022vgo}, as a tool to image nuclear ground states  \cite{STAR:2024eky}. For the spherical baseline, the elliptic flow ratio is below unity. This is due to  the larger mass number of $^{20}$Ne, which reduces initial-state fluctuations with respect to collisions involving $^{16}$O nuclei. Indeed, in absence of nuclear deformation corrections, a larger elliptic flow should be observed in Pb+O collisions.

Figures~\ref{fig:vnratio}(c)-(d) show our predictions for the rms triangular flow, $v_3\{2\}$, in both systems. The predictions seem diametrically opposite to those obtained for the elliptic flow: $v_3\{2\}$ depends strongly on the centrality percentile in Fig.~\ref{fig:vnratio}(c), while Fig.~\ref{fig:vnratio}(d) indicates little difference between Pb+Ne and Pb+O systems. This in agreement with the predictions obtained for the ratio taken between symmetric Ne+Ne and O+O collisions \cite{Giacalone:2024luz}.

Moving on to the quadrangular flow ($n=4$), the same patterns seen in the case of elliptic flow are recovered. In Fig.~\ref{fig:vnratio}(e), $v_4\{2\}$ shows little variation with centrality, though the ratio of quadrangular flow coefficients in Fig.~\ref{fig:vnratio}(f) is enhanced by as much as 20\% in central Pb+Ne collisions compared to Pb+O. As explicitly demonstrated in the SM, one can identify the origin of such behavior in the non-linear mode coupling between the elliptic and quadrangular flow vectors \cite{Gardim:2011xv,Teaney:2012ke,Yan:2015jma,Qian:2016fpi}.  Therefore, while the shape of $^{20}$Ne directly impacts the magnitude of $v_2$, it impacts the $v_4$ indirectly via nonlinear coupling. This extra layer of complexity leads to an even more stringent test of the hydrodynamic model. It should be investigated in experiments, and tested as well within a framework that is not hydrodynamic, such as transport \cite{Jia:2022qrq}.

\begin{figure}
      \includegraphics[width=0.8\columnwidth]{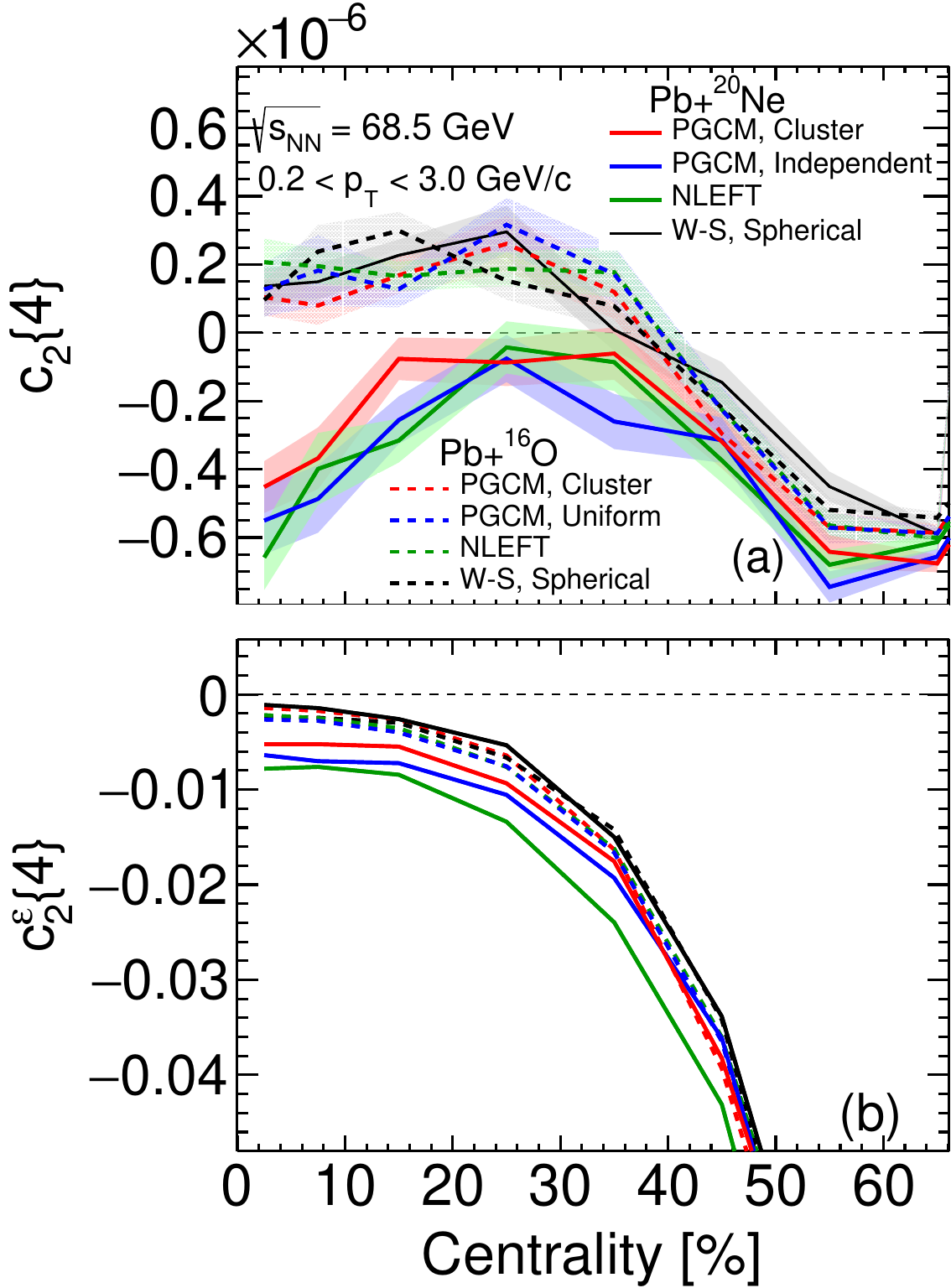}
    \caption{ Centrality dependence of $c_2\{4\}$ computed both from the final-state hadron distributions (upper) and from the initial-state eccentricity distribution (lower) in Pb+Ne collisions (solid lines) and Pb+O collisions (dashed lines) at $\sqrt{s_{\rm NN}}=$ 68.5 GeV. Different line colors correspond to different nuclear structure inputs. Errors are statistical only and of the same size as the shown lines when not visible.} 
    \label{fig:c24}
\end{figure}

%% Figure 4
Before concluding, we analyze a standard probe of hydrodynamic behavior in high-energy collisions, the  fourth-order cumulant of the elliptic flow distribution:
\begin{equation}
\label{eq:c24}
    c_2\{4\} = \langle v_2^4 \rangle - 2 \langle v_2^2 \rangle ^2,
\end{equation}
where averages are again over events at a given centrality. As we explain in the SM,  this quantity measures the kurtosis of the  elliptic flow vector distribution at a given centrality \cite{Abbasi:2017ajp,Bhalerao:2018anl}, and can have either negative or positive sign. The centrality dependence of $c_2\{4\}$ is shown in Fig.~\ref{fig:c24}(a). Surprisingly, in central collisions we predict:
\begin{equation}
\label{eq:c24pred}
    c_2\{4\}_{\rm PbNe} < 0 < c_2\{4\}_{\rm PbO}.
\end{equation}
To get some understanding, we calculate in Fig.~\ref{fig:c24}(b) the same quantity with $v_2$ in Eq.~(\ref{eq:c24}) replaced by the initial-state ellipticity, $\varepsilon_2$. The sign of the resulting cumulant is negative, and its magnitude is much larger in Pb+Ne collisions than in Pb+O collisions, in agreement with previous studies with deformed nuclei \cite{Giacalone:2018apa,Mehrabpour:2023ign}.  The hydrodynamic expansion adds a positive correction to  the value of $c_2\{4\}$ in a given centrality, changing even the sign of the cumulant in Pb+O collisions, or when a spherical neon is used. Similar results are found for $p$-$p$ collisions \cite{Zhao:2017rgg,Zhao:2020pty}. However, the effect of the   $^{20}$Ne shape is opposite: the impact of the deformation reduces the kurtosis, causing $c_2\{4\}$ in Pb+Ne collisions to preserve its negative sign after the hydrodynamic evolution. This interplay between the deformation effect and hydrodynamic response provides a nontrivial probe of the dynamics of the collisions, and so we urge  the experiments to verify this result.  

Note that the parameters of the hydrodynamic model used here do not yet result from a Bayesian analysis of the RHIC BES data. Though unlikely, we can not exclude \textit{a priori} that there may exist  a plausible set of parameters (likely, initial-state parameters) for which $c_2^\varepsilon\{4\}$ has a different sign. In addition, the choice of the variable  used to the define the collision centrality affects the sign of the cumulant \cite{Zhou:2018fxx,ATLAS:2019peb}.  Systematic analyses to address these points are costly for 3+1D simulations, and outside the scope of present manuscript, but should be carried out for future comparisons with the experimental results. 

%\noindent {\it 4. Conclusions and Outlook. }
In summary, we have unveiled the great opportunities offered by studies of the anisotropic flow in fixed-target collisions at the LHCb detector. Our predictions may be amenable to experimental verification already with the next LHC ion run. The availability of the bowling-pin nucleus $^{20}$Ne in SMOG2 enables one to isolate strong effects of the hydrodynamic response. If the dramatic enhancements of $v_2$ and $v_4$ relative to the Pb+O baseline will be confirmed by the LHCb collaboration, they will hint at the validity of a QGP description. This will pave the way to quantitative characterizations of the matter formed in these experiments via future Bayesian analyses in hydrodynamic calculations. Moreover, it would be insightful to compare these hydrodynamic results to those obtained from alternative dynamical approaches, such as transport \cite{Lin:2021mdn}, based on the same \textit{ab initio} nuclear input. This will permit us to more precisely pin down the microscopic mechanisms underlying the three-dimensional flow produced in the ion-gas collisions, with good control over the interaction geometry.

The  predictions of this manuscript can be improved in several directions. The impact of short-range two-particle correlations (\textit{non flow}) is washed out in our simulations due to the oversampling of the UrQMD events, which can be relaxed in future studies. In addition, a more detailed analysis of the systematics of the theoretical model will elucidate whether   theoretical uncertainties cancel when normalizing Pb+Ne observables with those coming from another collision species \cite{Giacalone:2024luz}. 

Additional observables can be studied to fully exploit the LHCb data. Focusing on $\eta$-differential quantities will help constrain models of longitudinal fluctuations, which are  under intense study in the community \cite{Soeder:2023vdn,McDonald:2023qwc,Garcia-Montero:2023gex,Ipp:2024ykh,Zhang:2024bcb}. In addition, probing hydrodynamic behavior via shape-induced modifications of the collective flow can be done as well through the mean transverse momentum, $\langle p_T \rangle$, of final-state hadrons,  by measuring its event-by-event fluctuations and the $v_2^2$-$\langle p_T \rangle$ correlation \cite{Giacalone:2019pca,Giacalone:2020awm,Jia:2021wbq,Jia:2021tzt,Bally:2021qys,Jia:2021qyu,ALICE:2021gxt,ATLAS:2022dov,Fortier:2023xxy,Nielsen:2023znu,STAR:2024eky,Wang:2024vjf,Fortier:2024yxs}.

Finally, other species are available in SMOG2, notably, nitrogen (N), argon (Ar), krypton (Kr), and xenon (Xe). Their clustered and deformed geometries will have to be elucidated via future \textit{ab initio} computations of nuclear structure based on high-resolution chiral EFT interactions \cite{Epelbaum:2008ga,Piarulli:2019cqu}, which have only recently been pushed to the description of medium-mass deformed nuclei \cite{Elhatisari:2022qfr,Frosini:2024ajq,Sun:2024iht,Hu:2024pee}. We are thus looking forward to an exciting program of cross-disciplinary QCD studies centered around the unique high-energy fixed-target mode of the LHC.

The gas-injection system of LHCb has been so far used with a natural neon source. Abundances of stable isotopes in the gas are 90.48\% for $^{20}$Ne, 0.27\% for $^{21}$Ne, and 9.25\% for $^{22}$Ne \cite{Meija2016a}. 
We recommend a purified $^{20}$Ne sample for use in the experimental campaigns for a transparent interpretation of the physics results. This is especially important in the limit of ultra-central collisions with the largest numbers of participant nucleons.

\bigskip
We thank the LHCb Collaboration for sharing with us the data shown in Fig.~1. We thank in particular Giacomo Graziani and the participants of the program+workshop \textit{``Exploring Nuclear Physics Across Energy Scales 2024''} for useful discussions. We thank Mubarak Alqahtani for a careful reading of the manuscript. 
% Giacalone
G.G.~is funded by the Deutsche Forschungsgemeinschaft (DFG, German Research Foundation) – Project-ID 273811115 – SFB 1225 ISOQUANT, and under Germany's Excellence Strategy EXC2181/1-390900948 (the Heidelberg STRUCTURES Excellence Cluster).
% Shen (Chun), Schenke
This material is based upon work supported by the U.S. Department of Energy, Office of Science, Office of Nuclear Physics, under DOE Contract No.~DE-SC0012704 (B.P.S.) and Award No.~DE-SC0021969 (C.S.), and within the framework of the Saturated Glue (SURGE) Topical Theory Collaboration. C.S. acknowledges a DOE Office of Science Early Career Award.
% Zhao
 W.B.Z. is supported by DOE under Contract No. DE-AC02-05CH11231, by NSF under Grant No. OAC-2004571 within the X-SCAPE Collaboration, and within the framework of the SURGE Topical Theory Collaboration. This research was done using resources provided by the Open Science Grid (OSG)~\cite{Pordes:2007zzb, Sfiligoi:2009cct}, which is supported by the National Science Foundation award \#2030508.
 % UGM
This work is supported in part by the European Research Council (ERC) under the European Union's Horizon 2020 research and innovation programme (ERC AdG EXOTIC, grant agreement No. 101018170), by DFG and NSFC through funds provided to the Sino-German CRC 110 ``Symmetries and the Emergence of Structure in QCD" (NSFC Grant No.~11621131001, DFG Grant No.~TRR110). The work of UGM was supported in part by the CAS President's International Fellowship Initiative (PIFI) (Grant No.~2018DM0034).
% Roth
R.R.~is supported by the Deutsche Forschungsgemeinschaft (DFG, German Research Foundation) – Projektnummer 279384907 – SFB 1245.
% Rodriguez
T.R.R.~is supported by the Spanish MCIU (PID2021- 127890NB-I00).
% Bally
The PGCM calculations were performed using HPC resources from GENCI-TGCC (Contracts No.\ A0130513012 and A0150513012) and CCRT (TOPAZE supercomputer). 
% NLEFT
The NLEFT code development and production calculations utilized the following computational resources: the Gauss
Centre for Supercomputing e.V. (www.gauss-centre.eu) for computing time on the GCS Supercomputer JUWELS at J{\"u}lich
Supercomputing Centre (JSC) and special GPU time allocated on JURECA-DC;
the Oak Ridge Leadership Computing Facility through the INCITE award
``Ab-initio nuclear structure and nuclear reactions''; and the TUBITAK ULAKBIM High Performance and Grid Computing Center (TRUBA resources). 
% Noronha-Hostler
J.N.H. acknowledges financial support from the US-DOE Nuclear Science Grant No. DE-SC0023861 and within the framework of the Saturated Glue (SURGE) Topical Theory.

\section{SUPPLEMENTAL MATERIAL}

 %%%%%%%%%%%%%%%%%%%%%%%%%%%%%%%%%%%%%%%%%%%%%%%%%%%%%%%%%%%%%%%%%%
  %%%%%%%%%%%%%%%%%%%%%%%%%%%%%%%%%%%%%%%%%%%%%%%%%%%%%%%%%%%%%%%%%%
   %%%%%%%%%%%%%%%%%%%%%%%%%%%%%%%%%%%%%%%%%%%%%%%%%%%%%%%%%%%%%%%%%%
\subsection{3D Glauber Monte Carlo initial conditions}

\label{app:A}

The 3D Monte-Carlo Glauber model is a dynamical event-by-event initial condition \cite{Shen:2017bsr,Shen:2022oyg}, that provides the space-time and momentum distributions of the initial energy-momentum tensor and net baryon charge current, employing a classical string deceleration model \cite{Bialas:2016epd,Shen:2017bsr}. The transverse positions of valence quarks and the soft partonic cloud in wounded nucleons are sampled from a 2D Gaussian. After the collision, the deposited energy density distribution and baryon charge have Gaussian profiles in the transverse plane. The average rapidity loss function of the valence quarks and the soft partonic cloud with their incoming rapidity $y_\mathrm{init}$ in the collision pair rest frame is parameterized as~\cite{Shen:2018pty,Shen:2022oyg},

\begin{equation}
    \langle y_\mathrm{loss} \rangle (y_\mathrm{init}) = A y_\mathrm{init}^{\alpha_2} [\tanh(y_\mathrm{init})]^{\alpha_1 - \alpha_2}.
    \label{eq:ylossMean}
\end{equation}

The parameters are fixed by fitting the  measured pseudo-rapidity distributions of charged hadrons for Au+Au and light-ion collisions \cite{Shen:2018pty,Shen:2022oyg,Zhao:2022ugy,Zhao:2022ayk}.  The detailed implementation of this initial condition model is discussed in Ref.~\cite{Shen:2022oyg}. The pre-hydrodynamic flow is included by a finite transverse initial velocity~\cite{Zhao:2022ugy}.  Finally, the produced strings from individual nucleon-nucleon collisions are  treated as dynamical source terms for the hydrodynamic evolution~\cite{Okai:2017ofp, Shen:2017ruz,Shen:2017bsr,Shen:2022oyg}.

 %%%%%%%%%%%%%%%%%%%%%%%%%%%%%%%%%%%%%%%%%%%%%%%%%%%%%%%%%%%%%%%%%%
  %%%%%%%%%%%%%%%%%%%%%%%%%%%%%%%%%%%%%%%%%%%%%%%%%%%%%%%%%%%%%%%%%%
   %%%%%%%%%%%%%%%%%%%%%%%%%%%%%%%%%%%%%%%%%%%%%%%%%%%%%%%%%%%%%%%%%%
\section{Cumulants of flow fluctuations in the lab frame}

\label{app:B}

We recall here the definition and the physical meaning of the cumulants of anisotropic flow fluctuations, $v_n\{2\}$ and $c_2\{4\}$, analyzed in this manuscript.

We start with the harmonic decomposition of the measured charged-hadron spectrum,  
\begin{equation}
\label{eq:fourier}
    \frac{dN}{d\phi_p} = \sum_{n=-\infty}^{+\infty} \textbf{v}_n e^{in\phi_p},  \hspace{20pt}\textbf{v}_{-n}=\textbf{v}_n^*,
\end{equation}
 where $\phi_p$ is the azimuthal angle of transverse momentum and the set of complex coefficients $\textbf{v}_n$ define the anisotropic flow. Note that $\textbf{v}_n$ comes with a real and an imaginary part  
\begin{equation}
\textbf{v}_n = (v_x, v_y)    
\end{equation}
which depends on the choice of the coordinate frame. Typically, $x$ and $y$ label either the transverse plane in the laboratory frame or the frame of the reaction plane with $x$ chosen along the impact parameter direction. 

Observables constructed from the anisotropic flow in heavy-ion collisions are measured as statistical averages involving azimuthal angles, which are taken over collision events, typically at a fixed final-state multiplicity (centrality). In the sample of events,  the flow vector, $\textbf{v}_n$, has some underlying distribution:
\begin{equation}
    P(\textbf{v}_n) = P(v_x,v_y), \hspace{30pt} v_n = \sqrt{v_x^2 + v_y^2}.
\end{equation}
Experimentally, the aim is to characterize such probability distribution in a given centrality class, which yields detailed information about the evolution and the properties of the produced matter.

Cumulants fully characterize a probability distribution and are particularly useful in this context.  The cumulant generating function of the distribution of $\textbf{v}_n$ is defined by:
\begin{equation}
\label{eq:Gk}
 \ln G({\bf k}) = \ln \int d\textbf{v}_n P(\textbf{v}_n) e^{{\bf k} \cdot \textbf{v}_n}  =  \ln \bigl  \langle e^{{\bf k} \cdot \textbf{v}_n} \bigr \rangle,
\end{equation}
from which the cumulant of order $m$ is derived as $\partial^{(m)}_{\bf k} \ln G({\bf k}) |_{{\bf k}=0}$ . In nuclear collisions, while the magnitude of the impact parameter can be controlled from the number of final-state particles, its orientation is always random. Therefore, experiments only access the angle-averaged information about the $\textbf{v}_n$ harmonics, i.e., moments of the magnitude, $v_n$, but can not reconstruct information about the individual $v_{x,y}$ components. In practice, we write the vector ${\bf k}=(k \cos\varphi, k \sin \varphi)$, and we are forced to average over  the angle in Eq.~(\ref{eq:Gk}), which yields:
\begin{equation}
\label{eq:average}
\ln G(k) = \ln \biggl \langle \int_0^{2\pi} \frac{d\varphi}{2\pi} e^{\textbf{k}\cdot \textbf{v}_n } \biggr \rangle = \ln \bigl \langle I_0(kv_n) \bigr \rangle. 
\end{equation}
Information is lost and we end up with a modified Bessel function, $I_0$, which is an even function. The  cumulants of the angle-averaged distribution of the   $\textbf{v}_n$ vector, whose standard notation is $v_n\{m\}^m$, are obtained from the equality:
\begin{equation}
\label{eq:cumull}
  \ln \bigl \langle I_0(kv_n) \bigr \rangle = \sum_{m=2}^{\infty} c_m k^m v_n\{m\}^m,
\end{equation}
where $c_m=0$ for odd $m$, while  $c_2=1/4$, $c_4=-1/64$, $c_6=1/576$ \cite{Bhalerao:2018anl}. Expanding the left-hand side of the previous equation, 
\begin{equation}
\label{eq:bessel}
\ln \bigl \langle I_0(kv_n) \bigr \rangle = \ln \biggl ( 1 + \frac{\langle v_n^2 \rangle k^2}{4} + \frac{\langle v_n^4 \rangle k^4}{64}  + \frac{\langle v_n^6 \rangle k^6}{2304} + \ldots \biggr),
\end{equation}
expanding this logarithm in powers of $k$, and matching to the right-hand side of Eq.~(\ref{eq:cumull}), one obtains:
\begin{align}
\label{eq:cumulants}
\nonumber    v_n\{2\}^2 &= \langle v_n^2 \rangle, \\
\nonumber    v_n\{4\}^4 &= 2\langle v_n^2 \rangle^2 - \langle v_n^4 \rangle, \\
    v_n\{6\}^6 &= \frac{1}{4} \bigl ( \langle v_n^6 \rangle - 9 \langle v_n^4 \rangle \langle v_n^2 \rangle + 12 \langle v_n^2 \rangle^3  \bigr),
\end{align}
where averages are over events at a given centrality. The cumulants are thus expressed in terms of even moments of the distribution of $v_n$, of the form $\langle v_n^{2k} \rangle$. These moments represent the quantities that can be measured in the experiments from established multi-particle correlation techniques \cite{Ollitrault:2023wjk}.

As pointed out in Ref.~\cite{Giacalone:2016eyu}, it is insightful to relate the observable quantities on the left-hand side of Eq.~(\ref{eq:cumulants}) to an expansion involving the (non-measurable) cumulants of the individual $x$ and $y$ components of $\textbf{v}_n$. Up to fourth order, the cumulants of the projections are:
\begin{align}
\nonumber\kappa_{10}&=\langle v_x\rangle, \hspace{10pt}\kappa_{01}=\langle v_y \rangle, \\
\nonumber \kappa_{20}&=\left\langle( v_x-\kappa_{10})^2\right\rangle, \hspace{10pt} \kappa_{02}=\left\langle( v_y-\kappa_{01})^2\right\rangle,\\
\nonumber\kappa_{30}&=\left\langle( v_x-\kappa_{10})^3\right\rangle, \hspace{10pt} \nonumber\kappa_{03}=\left\langle( v_y-\kappa_{01})^3\right\rangle, \\
\nonumber    \kappa_{12} &= \langle (v_x-\kappa_{10}) (v_y-\kappa_{01})^2 \rangle  , \\
\nonumber \kappa_{21} &= \langle  (v_x-\kappa_{10})^2 (v_y-\kappa_{01}) \rangle  ,  \\
\nonumber \kappa_{40}&=\left\langle( v_x-\kappa_{10})^4\right\rangle-3\kappa_{20}^2,  \\
\nonumber \kappa_{04}&=\left\langle( v_y-\kappa_{01})^4\right\rangle-3\kappa_{02}^2,  \\
 \kappa_{22} &= \langle (v_x-\kappa_{10})^2 (v_y-\kappa_{01})^2 \rangle -  \kappa_{20}\kappa_{02}. 
\end{align}
If $x$ and $y$ parametrize the transverse plane in the lab frame, the odd moments of the $v_{x}$ and $v_{y}$ distributions vanish (for an ideal detector), meaning that we are only left with non-zero values for $\kappa_{20}$, $\kappa_{02}$, $\kappa_{40}$, $\kappa_{04}$, $\kappa_{22}$. A little algebra \cite{Giacalone:2016eyu,Abbasi:2017ajp,Bhalerao:2018anl} shows that the first two cumulants in Eq.~(\ref{eq:cumulants}) can consequently be written as:
\begin{align}
  \label{eq:exact}
 v_n\{2\}^2&=\kappa_{20}+\kappa_{02}, \\
 v_n\{4\}^4&= - (\kappa_{04}+\kappa_{40}+2\kappa_{22}). 
\end{align}
This implies the following. The second-order cumulant $v_2\{2\}^2$ isolates the variance of the two-dimensional distribution. It tells us how a change in the mean square value of the flow vector magnitude, $\langle v_n^2 \rangle$, is  associated with an increase in the variances of its $x$ and $y$ components. Concerning the fourth-order cumulant, conventionally denoted by \footnote{The sign comes from the negative sign of the $k^4$ term in the expansion of Eq.~(\ref{eq:bessel}).}
\begin{equation}
 c_2\{4\} \equiv -v_2\{4\}^4  , 
\end{equation}
it isolates instead the kurtosis correction, and vanishes if $P(v_x,v_y)$ is a two-dimensional Gaussian. 

We understand then the impact of the shape of $^{20}$Ne on the elliptic flow vector, $\textbf{v}_2$:
\begin{itemize}
    \item The increase of $v_2\{2\}$ induced by the deformation of the nucleus is due to an isotropic increase in spread of the two-dimensional $\textbf{v}_2$ distribution (see also \cite{Jia:2022qgl}).
    \item  The increase in $c_2\{4\}$ implies instead that larger elliptic flow fluctuations lead as well to enhanced non-Gaussian tails. The kurtosis comes with a positive sign in the expression of $c_2\{4\}$, therefore, nuclear deformation reduces the kurtosis leading to a platykurtic distribution (tails thinner than a Gaussian)  at a given centrality.  
\end{itemize}

 %%%%%%%%%%%%%%%%%%%%%%%%%%%%%%%%%%%%%%%%%%%%%%%%%%%%%%%%%%%%%%%%%%
  %%%%%%%%%%%%%%%%%%%%%%%%%%%%%%%%%%%%%%%%%%%%%%%%%%%%%%%%%%%%%%%%%%
   %%%%%%%%%%%%%%%%%%%%%%%%%%%%%%%%%%%%%%%%%%%%%%%%%%%%%%%%%%%%%%%%%%
\section{Understanding the enhancement of $v_4$}

Here we explain the enhancement of $v_4\{2\}$ in central Pb+Ne collisions relative to central Pb+O collisions observed in Fig.~\ref{fig:vnratio}.  We recall that the quadrangular flow vector, $\textbf{v}_4=v_4e^{i4\phi_4}$, can by virtue of symmetry be decomposed as follows \cite{Yan:2015jma},
\begin{equation}
    \textbf{v}_4 = \textbf{v}_{4L} + \chi_{4} \textbf{v}_2^2,
\end{equation}
where $\chi_{4}$ is a coefficient that quantifies the strength of the nonlinear coupling to the squared elliptic flow vector, $\textbf{v}_2=v_2e^{i2\phi_2}$, while $\textbf{v}_{4L}$ is defined as the vector  statistically uncorrelated with $\textbf{v}_2$ in the considered centrality class, that is,
\begin{equation}
    \langle \textbf{v}_{4L} \textbf{v}_2^{2*} \rangle = 0.
\end{equation}
With these definitions, one obtains that the mean squared quadrangular flow is the sum of two contributions that add in quadrature,
\begin{equation}
    \langle v_4^2 \rangle = \langle v_{4L}^2 \rangle + \chi_4^2 \langle v_2^4 \rangle,
\end{equation}
where $v_4=|\textbf{v}_4|$, $v_2=|\textbf{v}_2|$, $v_{4L}=|\textbf{v}_{4L}|$. 

\begin{figure}
      \includegraphics[width=0.95\columnwidth]{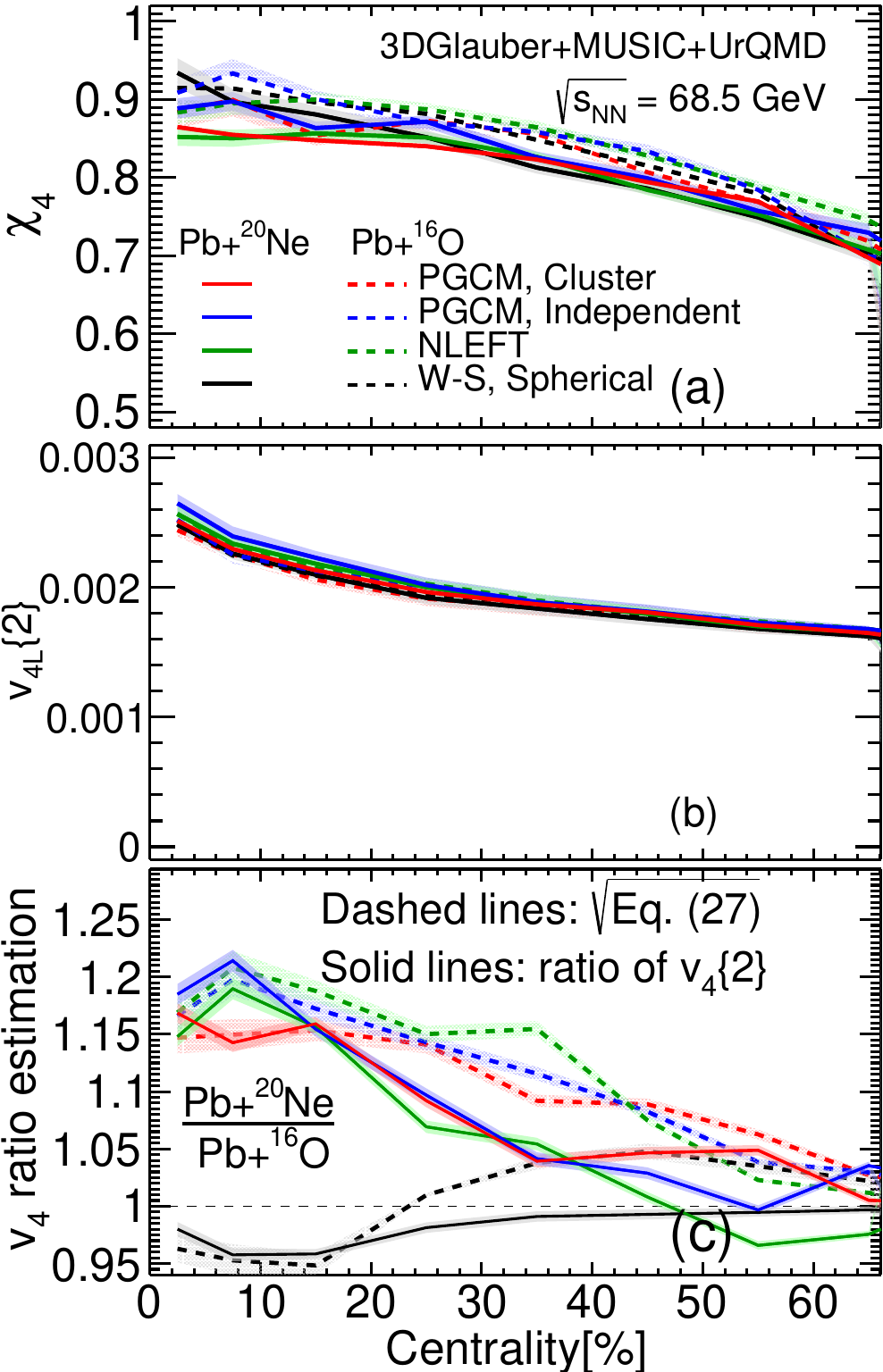}
    \caption{Centrality dependence of the non-linear coupling coefficient $\chi_{4}$ (a), of the \textit{linear} flow mode $v_{\rm 4L}\{2\}$ (b), and of the $v_4$ ratio (c) in Pb+Ne and Pb+O collisions at $\sqrt{s_{\rm NN}}=$ 68.5 GeV. The dashed line in panel (c) corresponds to the estimate of Eq.~(\ref{eq:v4est}). Errors are statistical only and of the same size as the shown lines when not visible. } 
    \label{fig:v4ratio}
\end{figure}

Now we take the ratio between Pb+Ne and Pb+O collisions, i.e.,
\begin{equation}
    \frac{\langle v_4^2 \rangle_{\rm PbNe}}{\langle v_4^2 \rangle_{\rm PbO}} = \frac{\langle v_{\rm 4L}^2 \rangle_{\rm PbNe} + \chi_{4,\rm PbNe}^2 \langle v_2^4 \rangle_{\rm PbNe}}{\langle v_{\rm 4L}^2 \rangle_{\rm PbO} + \chi_{4,\rm PbO}^2 \langle v_2^4 \rangle_{\rm PbO}}.
\end{equation}
In central collisions, one expects the term coupling to elliptic flow to be sub-leading compared to the uncorrelated contribution, $\langle v_{\rm 4L}^2 \rangle \gg \chi_4^2 \langle v_2^4 \rangle$. This justifies a Taylor expansion of the previous expression:
\begin{align}
\nonumber    & \frac{\langle v_4^2 \rangle_{\rm PbNe}}{\langle v_4^2 \rangle_{\rm PbO}} = \\
  \label{eq:v4ratio}  &\frac{\langle v_{\rm 4L}^2 \rangle_{\rm PbNe}}{\langle v_{\rm 4L}^2 \rangle_{\rm PbO}} + \frac{\chi_{4,\rm PbNe}^2 \langle v_2^4 \rangle_{\rm PbNe}}{\langle v_{\rm 4L}^2 \rangle_{\rm PbO}} - \frac{\langle v_{\rm 4L}^2 \rangle_{\rm PbNe}}{(\langle v_{\rm 4L}^2 \rangle_{\rm PbO})^2} \chi_{4,\rm PbO}^2 \langle v_2^4 \rangle_{\rm PbO}.
\end{align}
Now, we show in Fig.~\ref{fig:v4ratio} hydrodynamic results for $\chi_{4}$ [panel (a)] and $ v_{\rm 4L}\{2\}\equiv\sqrt{\langle v_{\rm 4L}^2 \rangle}$ [panel (b)]. We see that both these quantities are rather universal and receive only percent-level corrections if one varies the nuclear species/structure.  For simplicity, let us then consider that 
\begin{align}
    \langle v_{\rm 4L}^2 \rangle_{\rm PbNe} = \langle v_{\rm 4L}^2 \rangle_{\rm PbO} &\equiv \langle v_{\rm 4L}^2 \rangle, \\
    \chi_{4, \rm PbNe} = \chi_{4, \rm PbO} &\equiv \chi_4.
\end{align}
With this, we can rewrite Eq.~(\ref{eq:v4ratio}) as follows:
\begin{equation} 
\label{eq:v4est}
\frac{v_4^2\{2\}_{\rm PbNe}}{v_4^2\{2\}_{\rm PbO}} \equiv \frac{\langle v_4^2 \rangle_{\rm PbNe}}{\langle v_4^2 \rangle_{\rm PbO}} = 1 + \chi_4^2 \frac{\langle v_2^4 \rangle_{\rm PbNe}}{\langle v_{\rm 4L}^2 \rangle} \left ( 1 - \frac{ \langle v_2^4 \rangle_{\rm PbO}   } { \langle v_2^4 \rangle_{\rm PbNe} }  \right ).
\end{equation}
This equality holds in central collisions if the departure from unity in the quadrangular flow ratio is due to the nonlinear coupling term to the elliptic flow.  Plugging in numbers from the 0-5\% class, we obtain the results shown in Fig.~\ref{fig:v4ratio}(c). Agreement between the $v_4\{2\}$ ratio and the prediction of Eq.~(\ref{eq:v4est}) is excellent in central collisions, confirming that it is the nonlinear coupling to elliptic flow that drives the nontrivial Pb+Ne to Pb+O ratio.

\bibliographystyle{JHEP-2modlong.bst}

\bibliography{refs}

\end{document}